\documentclass[useAMS,usenatbib]{mn2e}
\usepackage{graphicx}
\begin{document}

\title[A spectroscopic study of the Globular Cluster NGC~4147]{A spectroscopic study of the
  Globular Cluster NGC~4147}
\author[S. Villanova et al.]{S. Villanova$^{1}$\thanks{E-mail: svillanova@astro-udec.cl (SV)}, 
L. Monaco$^{2}$, C. Moni Bidin$^{3}$, and P. Assmann$^{4,5,1}$\\
$^{1}$Departamento de Astronomia, Casilla 160, Universidad de Concepci{\'o}n, Chile\\
$^{2}$Universidad Andres Bello, Departamento de Ciencias Fisicas, Republica 220, Santiago, Chile\\
$^{3}$Instituto de Astronomia, Universidad Catolica del Norte, Av. Angamos 0610 Antofagasta Chile\\
$^{4}$NAOC, Chinese Academy of Sciences, 20A Datun Rd., Chaoyang District, 100012, Beijing, China\\
$^{5}$Departamento de Astronomia, Universidad de Chile, Camino El Observatorio 1515, Las Condes, Santiago, Chile\\
}

\date{Accepted --. Received --; in original form --}

\pagerange{\pageref{firstpage}--\pageref{lastpage}} \pubyear{2015}

\maketitle

\label{firstpage}

\begin{abstract}
We present the abundance analysis for a sample of 18 red giant branch stars  in
the metal-poor globular cluster NGC~4147 based on medium and high resolution spectra.
This is the first extensive spectroscopic study of this cluster. 
We derive abundances of C, N, O, Na, Mg, Al, Si, Ca, Ti, Cr, Fe, Ni, Y, Ba,
and Eu. We find a metallicity of [Fe/H]=-1.84$\pm$0.02 and an
$\alpha$-enhancement of +0.38$\pm$0.05 (errors on the mean), 
typical of halo globular clusters in this
metallicity regime. A significant spread is observed in the abundances of light
elements C, N, O, Na, and Al. In particular we found a Na-O anti-correlation
and Na-Al correlation. The cluster contains only $\sim$15\% of stars
that belong to the first generation (Na-poor and O-rich). This implies that it
suffered a severe mass loss during its lifetime.
Its [Ca/Fe] and [Ti/Fe]  mean values agree better with the Galactic Halo
trend than with the trend of extragalactic environments at the cluster
metallicity. This possibly suggests that NGC~4147 is a genuine Galactic object at odd
with what claimed by some author that proposed the cluster to be member of the Sagittarius
dwarf galaxy. A anti-relation between the light \textit{s-}process element
Y and Na may also be present. 
\end{abstract}


\section{Introduction}

Galactic globular clusters (GGCs) are known to host star-to-star variations as
far as chemical abundances are concerned.  
More specifically, \citet{Ca09b} showed that all GGCs studied up to now have at least a
spread (or anti-correlation) in the content of their light-elements O and Na.
The only confirmed exception is Ruprecht 106, where \citet{Vi13} found that
stars share a homogenous chemical composition. 
This spread is probably due to the early evolution of each
cluster, when a second generation of stars (Na-rich and O-poor) was born from gas polluted by
ejecta of evolved stars of the first generation (Na-poor and O-rich). This is
the so called multiple-population phenomenon. 

\citet{Ca09b} showed also that most of the stars currently found in a GC belong to the second
generation ($\sim$60$\div$80\%). This is at odds with theory, which says
that first generation stars must have been much more numerous than what we
observe nowadays in order to produce enough ejecta to form the second \citep{De08}. 
This contradiction can be partially explained if we assume that ejecta were
collected preferentially in the center of the cluster due to the gravitational potential. Because of this
the second generation was formed in the center and was much less affected by
the violent relaxation and the gas expulsion phase during the
proto-cluster period \citep{Kh15} or by  Galactic tidal disruption than the first,
which lost most of its members \citep{Ca11}.

On the other hand \citet{Ca11} suggested the possibility of the existence of clusters 
that retained almost all the first generation and so only a small
fraction of the stars would belong to the second.
An example of such a cluster is Terzan 8, where \citet{Ca14} found that almost all
stars belong to the first generation.

Here we present an opposite case, the Globular Cluster (GC) NGC~4147. This object has an
Horizontal Branch (HB) that is mainly populated in the blue part. However it
shows also the presence of a red tail. According to \citet{Ca11}, this
indicates that almost all its stars should belong to the second generation,
with a fraction of first generation objects that should be very small,
probably smaller than all the other GGCs studied  up to now. In order to verify
this statement we obtained spectroscopic data in the Red Giant Branch (RGB)
region with the aim of measuring their light-element content.
We take advantage of these data also to perform a full chemical analysis of
the cluster both to study the chemical trend of its multiple-populations 
and to compare it with different environments. This is because
NGC~4147, based on the projection on the sky of the theoretical orbit of the
Sagittarius dwarf spheroidal galaxy computed by \citet{Ib98}, was suggested to be a
possible former member of the galaxy \citep{Be03a}, like NGC~5053 and NGC~5634, recently
studied by \citet{Sb15}. \citet{Be03b} also found evidence for the presence of Sgr
tidal stream stars in the background of NGC\,4147 using 2MASS data.

\citet{La10}, on the other hand, used the spatial and kinematic
data available for stars associated to the Sgr tidal stream to construct
numerical model of the tidal disruption of the galaxy. These authors
considered the association of NGC\,4147 to Sgr as still possible, but
with a low statistical confidence. 

In section 2 we describe data reduction and in section 3 the methodology we
used to obtain the chemical abundances. In section 4 we present our results
including a comparison with different environments (Galactic and
extragalactic). Finally in section 5 we give a summary of our findings.  

\begin{table*}
\caption{Parameters for the observed stars including FeI, NaI, CaI and BaII
  abundances. Reported errors are errors on the mean.}            
\label{t1}      
\centering                          
\begin{tabular}{lccccccccccccc}        
\hline\hline                 
ID & \scriptsize{RA} & \scriptsize{DEC} & \scriptsize{B} & \scriptsize{V} & \scriptsize{RV$_{\rm H}$} &
\scriptsize{T$_{\rm {eff}}$} & \scriptsize{log(g)} & \scriptsize{v$_{\rm t}$} & \scriptsize{[Fe/H]} & \scriptsize{[Na/Fe]} & \scriptsize{[Na/Fe]$_{NLTE}$}  & \scriptsize{[Ca/Fe]} & \scriptsize{[Ba/Fe]}\\    
\hline      
      & & &  & &  &GIRAFFE&  &  &  & & & & \\
D1    & 182.55908 & 18.53358 & 17.641 & 16.845 & 180.2 & 4910 & 2.17 & 1.38  & -1.77  & 0.48 &0.47 & 0.28& -0.22\\
D8    & 182.53150 & 18.51826 & 16.128 & 15.086 & 178.0 & 4480 & 1.23 & 1.62  & -1.81  & 0.39 &0.52 & 0.44& -0.23\\
S268  & 182.50658 & 18.52611 & 17.355 & 16.542 & 180.3 & 4840 & 2.02 & 1.42  & -1.80  & 0.47 &0.48 & 0.25& -0.41\\
S377  & 182.47029 & 18.55802 & 17.506 & 16.725 & 177.9 & 4880 & 2.11 & 1.39  & -1.86  & 0.73 &0.73 & 0.55& -0.04\\
S408  & 182.50379 & 18.55402 & 17.496 & 16.701 & 177.3 & 4880 & 2.10 & 1.40  & -1.81  & 0.54 &0.54 & 0.29& -0.31\\
S414  & 182.50837 & 18.53466 & 17.642 & 16.849 & 180.2 & 4910 & 2.18 & 1.38  & -1.80  & 0.56 &0.55 & 0.51& -0.34\\
S429  & 182.51487 & 18.53925 & 16.830 & 15.950 & 178.7 & 4710 & 1.71 & 1.49  & -1.90  & 0.70 &0.75 & 0.59& -\\
S432  & 182.51587 & 18.56322 & 16.756 & 15.777 & 178.7 & 4670 & 1.62 & 1.52  & -1.86  & 0.60 &0.67 & 0.52& -0.20\\
S445  & 182.51976 & 18.55226 & 16.267 & 15.296 & 181.0 & 4540 & 1.35 & 1.59  & -1.75  & 0.39 &0.50 & 0.30& -0.23\\
S688  & 182.53112 & 18.53497 & 16.619 & 15.688 & 182.1 & 4640 & 1.57 & 1.53  & -1.82  & 0.49 &0.57 & 0.34&  0.03\\
S690  & 182.53459 & 18.53544 & 16.556 & 15.629 & 177.6 & 4630 & 1.54 & 1.54  & -1.71  & 0.51 &0.59 & 0.33& -0.31\\
S692  & 182.53587 & 18.54886 & 17.532 & 16.737 & 181.3 & 4880 & 2.12 & 1.39  & -1.92  & 0.61 &0.61 & 0.58& -0.12\\
\hline
Cluster     &           &          &        &        &       &      &      &       & -1.79  & 0.48 &0.58 & 0.42&-0.22\\
Error       &           &          &        &        &       &      &      &       &  0.03  & 0.05 &0.03 & 0.04&0.04\\
\hline
      & & &  & &  &MIKE&  &  &  & & & & \\                                                                                               
D8    & 182.53150 & 18.51826 & 16.128 & 15.086 & 178.9 & 4430 & 0.80 & 1.70  & -1.84  &  0.56 & 0.46 & 0.38 & -0.19\\
S437  & 182.51697 & 18.54816 & 15.885 & 14.716 & 184.0 & 4300 & 0.45 & 1.80  & -1.88  & -0.03 &-0.11 & 0.39 & -0.22\\
S445  & 182.51976 & 18.55226 & 16.267 & 15.296 & 181.9 & 4460 & 0.75 & 1.59  & -1.86  &  0.69 & 0.59& 0.35 & -0.17\\
S454  & 182.52253 & 18.54305 & 16.257 & 15.266 & 177.8 & 4470 & 0.85 & 1.67  & -1.84  &  0.47 & 0.36& 0.45 & -0.15\\
S468  & 182.52524 & 18.54625 & 15.791 & 14.702 & 182.6 & 4300 & 0.22 & 1.79  & -1.91  &  0.18 & 0.10& 0.37 & -0.04\\
S690  & 182.53459 & 18.53544 & 16.556 & 15.629 & 179.2 & 4570 & 1.14 & 1.53  & -1.82  &  0.71 & 0.57&  -   & -0.05\\
\hline                            
Cluster     &           &          &        &        &       &      &      &       & -1.86  &  0.43 &0.33 & 0.39 &-0.14\\
Error       &           &          &        &        &       &      &      &       &  0.01  &  0.12 &0.11 & 0.02 & 0.03\\       
\hline
$\Delta$(GIR-MIK) & - & - & - & - & - & +63 & +0.48 & -0.03 & +0.08 & - & - & +0.00 & -0.12\\ 
\end{tabular}
\end{table*}

\begin{table*}
\caption{Chemical abundances of MIKE stars. The abundance for Ti is the
    mean of those obtained from the neutral and singly ionized
    species. Reported errors are errors on the mean.}            
\label{t2}      
\centering                          
\begin{tabular}{lcccccccccccc}        
\hline\hline                 
ID & \scriptsize{[CI/Fe]} & \scriptsize{[NI/Fe]} & \scriptsize{[OI/Fe]} &\scriptsize{[CNO/Fe]} & \scriptsize{[MgI/Fe]} & \scriptsize{[AlI/Fe]} & \scriptsize{[SiI/Fe]} & \scriptsize{[Ti/Fe]} & \scriptsize{[CrI/Fe]} & \scriptsize{[NiI/Fe]} & \scriptsize{[YII/Fe]} & \scriptsize{[EuII/Fe]}\\    
\hline      
D8    &  -0.72 & 0.86 & -0.13 &   0.07 &  0.38 & 0.95 & 0.44 &  0.30 & -0.16 &  0.03 & -0.22 & 0.45\\
S437  &  -0.24 & 0.21 &  0.30 &   0.21 &  0.47 & 0.25 & 0.47 &  0.31 & -0.11 & -0.06 & -0.01 & 0.38\\
S445  &  -0.64 & 0.75 & -0.27 &  -0.04 &  0.38 & 1.03 & 0.44 &  0.25 & -0.18 & -0.08 & -0.33 & 0.33\\
S454  &  -0.26 & 0.60 &  0.17 &   0.17 &  0.50 & 0.91 &  -   &  0.33 & -0.23 & -0.07 & -0.35 & 0.37\\
S468  &  -0.53 & 0.54 &  0.26 &   0.20 &  0.46 & 0.27 & 0.49 &  0.27 & -0.14 & -0.07 & -0.11 & 0.27\\
S690  &  -0.61 & 0.80 &  0.06 &   0.14 &  0.34 & 1.01 &  -   &  0.25 &   -   & -0.05 & -0.43 & 0.42\\
\hline        
Cluster     &  -0.50 & 0.63 &  0.07 &   0.12 &  0.42 & 0.74 & 0.46 & 0.28 & -0.16 & -0.05 & -0.24 & 0.37\\
Error       &   0.08 & 0.10 &  0.09 &   0.04 &  0.03 & 0.15 & 0.01 & 0.01 &  0.02 &  0.02 &  0.06 & 0.03\\          
\end{tabular}
\end{table*}

\begin{table*}
\caption{Estimated errors on abundances due to errors on atmospheric
parameters and to spectral noise compared with the observed errors for stars
\#D8. The last column gives the observed dispersion of MIKE data.}
\label{t3}      
\centering                          
\begin{tabular}{lcccccccccccccc}        
\hline\hline  
ID & $\Delta$T$_{\rm eff}$=50 K  & $\Delta$log(g)=0.10 & $\Delta$v$_{\rm t}$=0.05 km/s
& $\Delta$[Fe/H]=0.05 & S/N & $\sigma_{\rm tot}$ & $\sigma_{\rm obs}$\\
\hline
$\Delta$([C/Fe])  & 0.02 & 0.02 & 0.05 & 0.04 & 0.02 & 0.07 & 0.20$\pm$0.06\\
$\Delta$([N/Fe])  & 0.05 & 0.02 & 0.05 & 0.03 & 0.02 & 0.08 & 0.23$\pm$0.07\\
$\Delta$([O/Fe])  & 0.05 & 0.06 & 0.05 & 0.02 & 0.02 & 0.10 & 0.23$\pm$0.07\\
$\Delta$([Na/Fe]) & 0.02 & 0.01 & 0.00 & 0.01 & 0.05 & 0.06 & 0.30$\pm$0.09\\
$\Delta$([Mg/Fe]) & 0.03 & 0.01 & 0.00 & 0.01 & 0.05 & 0.06 & 0.06$\pm$0.02\\
$\Delta$([Al/Fe]) & 0.02 & 0.00 & 0.02 & 0.01 & 0.05 & 0.06 & 0.39$\pm$0.11\\
$\Delta$([Si/Fe]) & 0.05 & 0.01 & 0.02 & 0.01 & 0.05 & 0.07 & 0.02$\pm$0.01\\
$\Delta$([Ca/Fe]) & 0.00 & 0.00 & 0.00 & 0.00 & 0.04 & 0.04 & 0.04$\pm$0.01\\
$\Delta$([Ti/Fe]) & 0.03 & 0.00 & 0.00 & 0.01 & 0.03 & 0.04 & 0.03$\pm$0.01\\
$\Delta$([Cr/Fe]) & 0.03 & 0.00 & 0.00 & 0.00 & 0.03 & 0.04 & 0.05$\pm$0.01\\
$\Delta$([Fe/H])  & 0.05 & 0.01 & 0.02 & 0.01 & 0.01 & 0.05 & 0.03$\pm$0.01\\
$\Delta$([Ni/Fe]) & 0.00 & 0.00 & 0.00 & 0.00 & 0.02 & 0.02 & 0.04$\pm$0.01\\
$\Delta$([Y/Fe])  & 0.07 & 0.04 & 0.03 & 0.00 & 0.04 & 0.09 & 0.16$\pm$0.05\\
$\Delta$([Ba/Fe]) & 0.03 & 0.05 & 0.03 & 0.03 & 0.01 & 0.07 & 0.07$\pm$0.02\\
$\Delta$([Eu/Fe]) & 0.05 & 0.03 & 0.03 & 0.02 & 0.02 & 0.07 & 0.04$\pm$0.01\\
\hline                                   
\end{tabular}
\end{table*}

\section{Observations, data reduction, and Abundance analysis}

Our dataset consists of medium and high resolution spectra collected at the
GIRAFFE (mounted at the VLT-UT2 telescope) and MIKE (mounted at the
Magellan-Clay telescope) spectrograph respectively \footnote{Observations
were taken with ESO telescopes at the La Silla-Paranal Observatory under programme ID
083.D-0530, and with LCO telescopes at the Magellan observatory under
programme ID CN2011B-039.}. Targets were selected
from the \citet{St05} photometry. Three targets were in common between the two datasets.

We observed with GIRAFFE a total of 17 RGB stars with magnitude between V=15.6 and V=18.0.
We used the set-up HR12, that gives a spectral coverage between 5820 and 6140
\AA\ with a resolution of R=18700. The signal-to-noise (S/N) is between 25 and 70 at 6000 \AA.
Data were reduced using the dedicated pipeline (see
http://www.eso.org/sci/software/pipelines/).
Data reduction includes bias subtraction, flat-field correction, wavelength calibration,
sky subtraction, and spectral rectification.

MIKE was used to observe 6 RGB stars with V magnitude between 14.7 and 15.6.
The spectrograph cover a wide spectral range, from the ultraviolet to the near
infrared, with a resolution of R=32000 (using a slit of 0.7 arcsec).
The S/N of the spectra is between 70 and 90 at 6000 \AA.
Data were reduced using IRAF \footnote{IRAF is distributed by the National
Optical Astronomy Observatory, which is operated by the Association of
Universities for Research in Astronomy, Inc., under cooperative agreement
with the National Science Foundation.}, including bias subtraction,
flat-field correction, wavelength calibration, scattered-light and sky
subtraction, and spectral rectification. 

\begin{figure}
 \includegraphics[width=80mm]{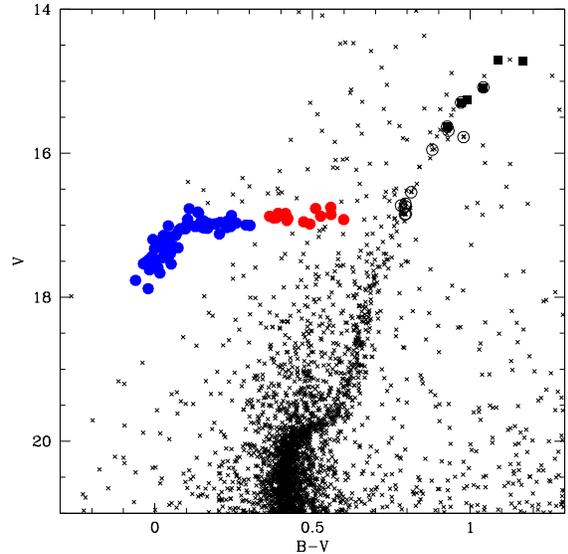}
 \caption{V vs. B-V CMD of NGC~4147. Open black circles indicate GIRAFFE
   targets, while filled black circles indicate MIKE targets. HB stars were
   divided in first generation (filled red circles) and second generation
   (filled blue circles) objects. See text for more details.}
 \label{f1}
\end{figure}

Radial velocities were measured by the {\it fxcor} package in IRAF,
using a synthetic spectrum as a template. Stars with very
different radial velocity with respect to the median value were rejected as
non-member. We rejected also two GIRAFFE targets with a radial velocity
similar to that of the cluster but with higher metallicities ([Fe/H]$\sim$-1.60). These
two stars have low S/N and we could use only one out of the three iron lines
available to measure [Fe/H] (see next section). So the disagreement in
metallicity is probably due to the
combination of the two factors.
We end up with a sample of 12 GIRAFFE and 6 MIKE targets.
The mean radial velocity we obtained is 179.9$\pm$0.5 km/s. \citet{Pe86} and
\citet{Pr88} give 183.5$\pm$1.3 and 183.2$\pm$0.9 instead. However the first
is based on one star only, while the second does not give any detail on the
radial velocity measurements they perform so we cannot suggest any reason for
this discrepancy.
If we compare the mean radial velocity obtained with the two instruments, we
get 179.4$\pm$0.5 for GIRAFFE, and 180.7$\pm$1.0 for MIKE (errors on the mean). The two values well
agree within 1$\sigma$.
Table~\ref{t1} lists the basic parameters of the retained stars:
ID (from \citealt{St05}), J2000.0 coordinates (RA \& DEC in degrees), B and V
magnitudes, heliocentric radial velocity RV$_{\rm H}$ (km/s),
T$_{\rm {eff}}$ (K), log(g), micro-turbulence velocity v$_{\rm t}$ (km/s),
plus [Fe/H], [Na/Fe], [Ca/Fe], and [Ba/Fe] abundances. The determination of the atmospheric parameters
and abundances is discussed in the next section. In Fig.~\ref{f1} we report,
on the top of the cluster color magnitude diagram (CMD), the MIKE targets as
black filled squares, and with open black circles the GIRAFFE targets.

\section{Abundance analysis}

Atmospheric parameters were obtained in the following way.  
First, T$_{\rm eff}$ was derived from the B-V color using the relation of
\citet{Ra05}. The reddening we adopted (E(B-V)=0.02) was
obtained from \citet[2010 edition]{Ha96}. 
Surface gravities (log(g)) were obtained from the canonical equation:
$$ \log\left(\frac{g}{g_{\odot}}\right) =
         \log\left(\frac{M}{M_{\odot}}\right)
         + 4 \log\left(\frac{T_{\rm{eff}}}{T_{\odot}}\right)
         - \log\left(\frac{L}{L_{\odot}}\right). $$
where the mass M/M$_{\odot}$ was assumed to be 0.8 M$_{\odot}$, and the
luminosity L/L$_{\odot}$ was obtained from the absolute magnitude M$_{\rm V}$
assuming an apparent distance modulus of (m-M)$_{\rm V}$=16.49 \citep{Ha96}. The
bolometric correction (BC) was derived by adopting the relation 
BC-T$_{\rm eff}$ from \citet{Al99}.
Finally, micro-turbulence velocity (v$_{\rm t}$) was obtained from the
relation of \citet{Ma08}.\\
Atmospheric models were calculated using ATLAS9 \citep{Ku70}
assuming our estimations of T$_{\rm eff}$, log(g), and v$_{\rm t}$, and the
[Fe/H] value from \citet{Ha96}([Fe/H]=-1.80).\\ 
The Local Thermodynamic Equilibrium (LTE) program MOOG \citep{Sn73} was used
for the abundance analysis.
For MIKE data, T$_{\rm eff}$, log(g), and v$_{\rm t}$ were re-adjusted and new 
atmospheric models calculated in an interactive way in order to remove trends 
in excitation potential and EQW versus abundance for T$_{\rm eff}$ and v$_{\rm t}$, respectively, 
and to satisfy the ionization equilibrium for log(g). 30$\div$40 FeI and
 8$\div$10 FeII (depending on the S/N of the spectrum) were used for the
latter purpose. The [Fe/H] value of the model was changed at each 
iteration according to the output of the abundance analysis. 

For GIRAFFE data we measured Fe, Na, Ca and Ba abundances.  Fe abundances were obtained 
from the equivalent width of the three iron lines at 5914, 5930, and 6065 \AA, while for Na we compared 
the strength of the NaD doublet at 589 nm with suitable synthetic spectra calculated 
for five different abundances. Finally Ca and Ba abundances were obtained
from the lines at 5857, 6102, and 5853 respectively using spectrosynthesis.
MIKE data allowed us to perform a more complete abundance 
analysis and we present here our measurements for C, N, O,  Na, Mg, Al, Si, Ti, Cr, Fe, Ni, Y, Ba, and Eu.
The line-list and the methodology we used are the same used in previous papers (e.g. \citealt{Vi13}),
so we refer to those articles for a detailed discussion about this point. 
Here we just underline that we took hyperfine splitting into account for Ba.
Na abundances were corrected for NLTE using the corrections provided by the INSPEC
\footnote{version 1.0 (http://inspect.coolstars19.com/index.php?n=Main.HomePage)} database.
After the NLTE correction, we found a small offset of 0.24 dex between Na abundances of the two datasets, that we
could estimate using the 3 stars in common (\#S445, \#S690, \#D8). This is
very likely due to some systematic error in the abundance determination
of the NaD doublet. This feature is saturated and then its abundance very sensitive to any
error in the flat-field or continuum normalization.
In order to remove it, we applied a correction of +0.24 dex to GIRAFFE NaD measurements.
Abundances for each target are reported in Tables~\ref{t1} and ~\ref{t2}, together with the mean values
and the error on the mean.

A detailed internal error analysis was performed by varying T$_{\rm eff}$, log(g), [Fe/H], and
v$_{\rm t}$ and redetermining abundances of star \#D8, assumed to represent
the entire MIKE sample. GIRAFFE data will not be used to check for
intrinsic spreads, so error analysis is not required for them. We just give in Tab.~\ref{t1}
the mean differences in T$_{\rm eff}$, log(g), v$_{\rm t}$, [Fe/H], [Ca/Fe],
and [Ba/Fe] with respect MIKE data based on the three stars in common. 
Parameters for stars \#D8 were varied by $\Delta$T$_{\rm eff}$=+50 K,
$\Delta$log(g)=+0.10, $\Delta$[Fe/H]=+0.05 dex, and $\Delta$v$_{\rm t}$=+0.05
km/s. This estimation of the internal errors for atmospheric parameters was
performed as in \citet{Ma08}.
Results are shown in Tab.~\ref{t3}, including the error due to the noise
of the spectra. This error was obtained for elements whose abundance was
obtained by EQWs, as the errors on the mean given by
MOOG, and for elements whose abundance was obtained by spectrum-synthesis, as
the error given by the fitting procedure. $\sigma_{\rm tot}$ is the
squared sum of the single errors, while $\sigma_{\rm obs}$ is the mean
observed dispersion of MIKE data.

\begin{figure}
 \includegraphics[width=80mm]{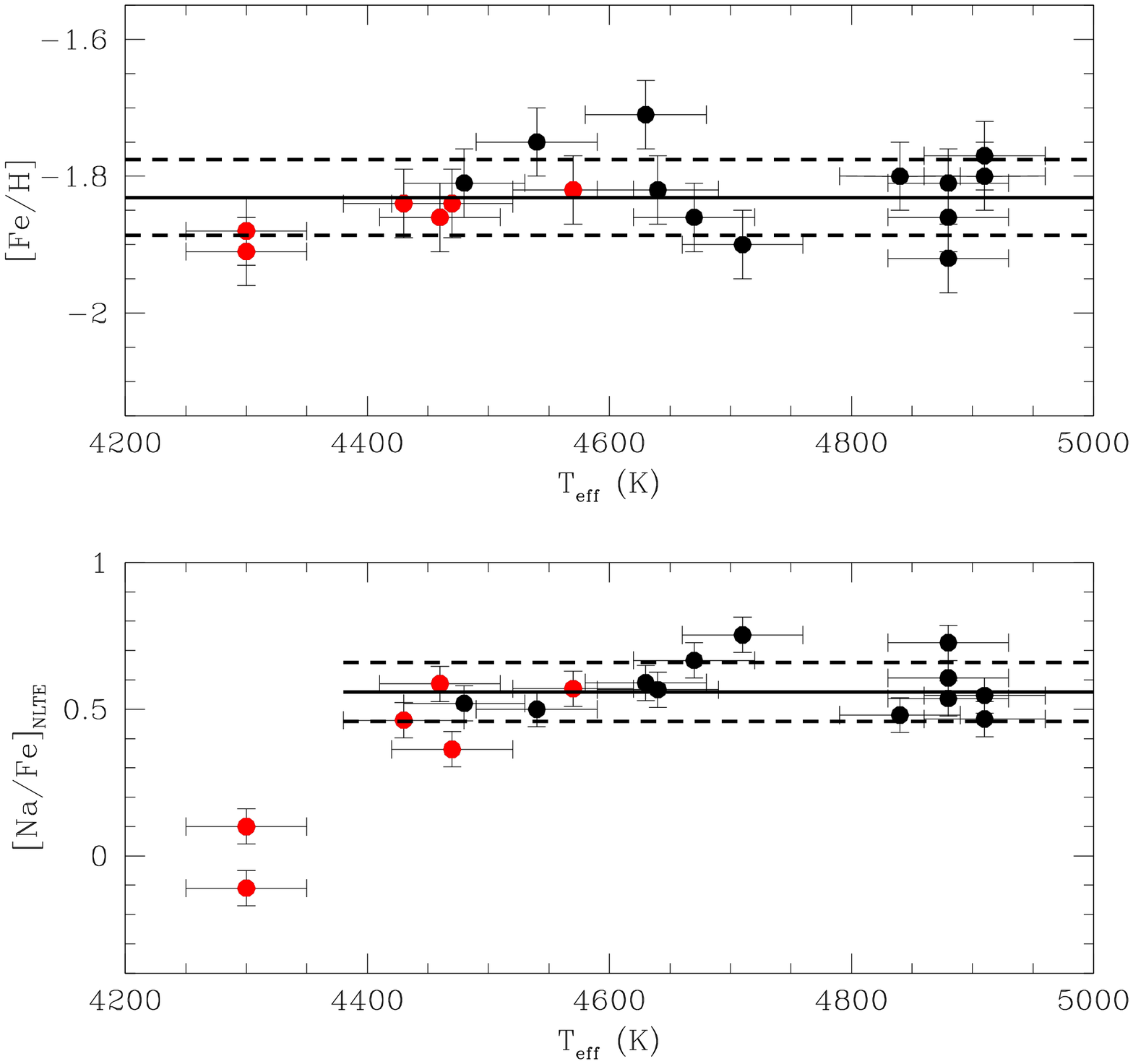}
 \caption{[Fe/H] vs. T$_{eff}$ (upper panel) and [Na/Fe] vs. T$_{eff}$ (lower
   panel) relations for our sample. GIRAFFE targets are reported as black
   circles, while MIKE targets are reported as red circles. Error adopted are
   those from Tab.~\ref{t3}. For both relations we
   indicated the mean value with a continuous black line and the
   $\pm$1$\sigma$ interval with two dashed lines. For the [Na/Fe]
   vs. T$_{eff}$ relation we considered only targets hotter than 4400 K. See
   text for more details.}
 \label{f1b}
\end{figure}

\begin{figure}
 \includegraphics[width=80mm]{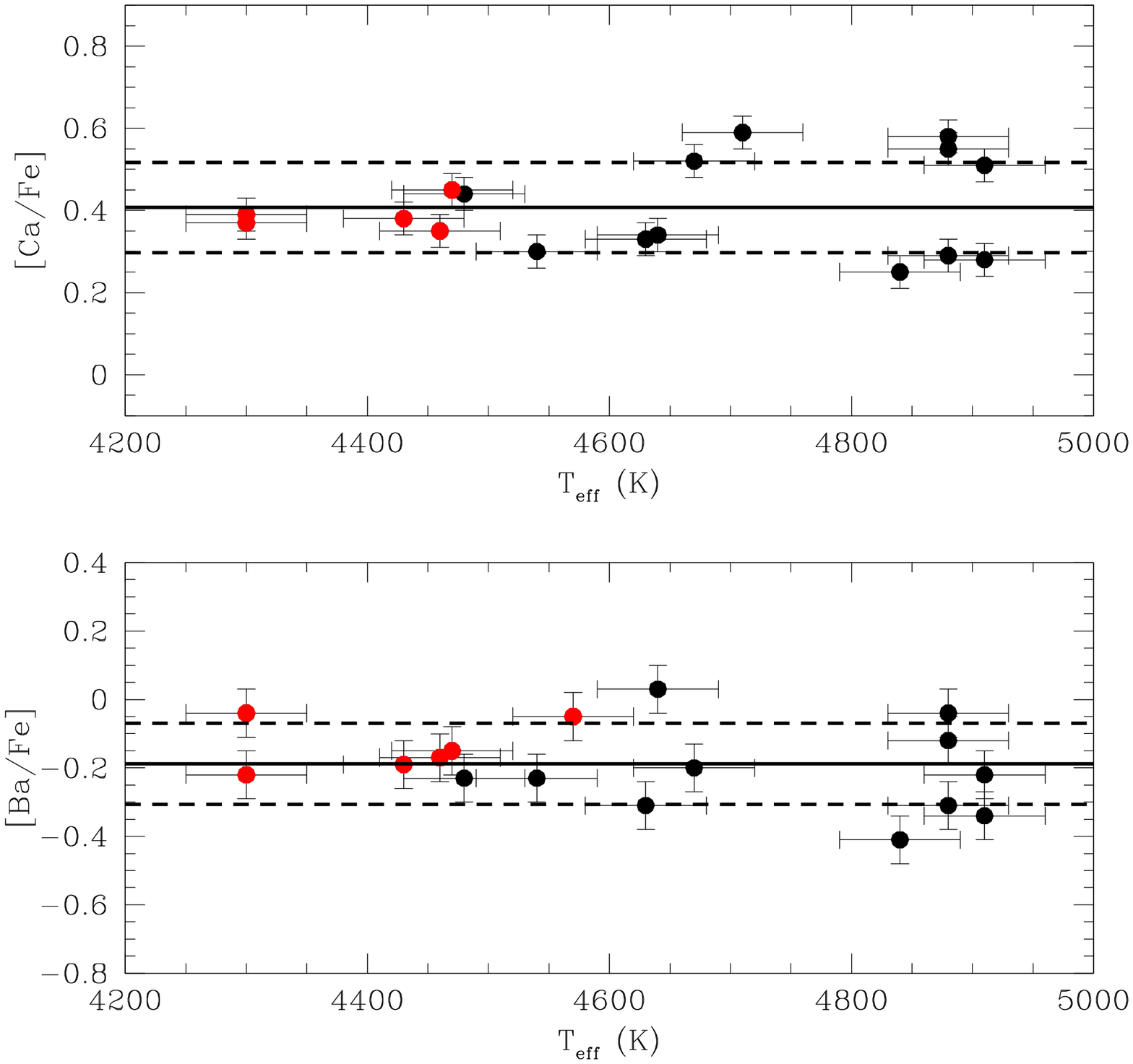}
 \caption{[Ca/Fe] vs. T$_{eff}$ (upper panel) and [Ba/Fe] vs. T$_{eff}$ (lower
   panel) relations for our sample. GIRAFFE targets are reported as black
   circles, while MIKE targets are reported as red circles. Error adopted are
   those from Tab.~\ref{t3}. For both relations we
   indicated the mean value with a continuous black line and the
   $\pm$1$\sigma$ interval with two dashed lines.}
 \label{f1c}
\end{figure}

\begin{figure}
 \includegraphics[width=80mm]{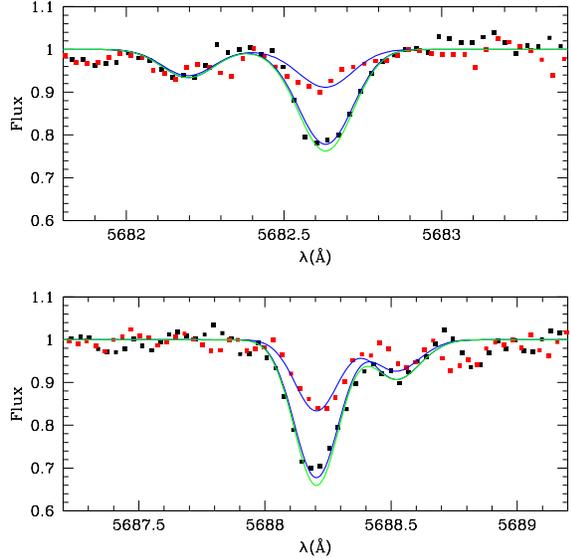}
 \caption{Spectra of the \#D8 (black squares) and \#S437 (red squares) targets
  around the Na double at 568 nm. Blue lines are the best fitting synthetic
  spectra with Na abundances taken from Tab.~\ref{t1}. The green line is a synthetic
  spectrum calculated for the atmospheric parameters of \#S437 but with the Na
  content of \#D8.}
 \label{f1d}
\end{figure}

As a cross check of our abundance analysis, we plot in Fig.~\ref{f1b} and Fig.~\ref{f1c} [Fe/H], [Na/Fe], [Ca/Fe], and [Ba/Fe]
vs. T$_{eff}$ for the entire sample. The temperature range covered by our
stars is about 600 K. As far as iron, calcium and barium are concern, a linear fit gives a slopes
with a significance below 1$\sigma$. We plot also the mean abundance for
each element and the $\pm$1$\sigma$ error. Again all stars are spread around
the mean value and no sign of trend is present. After this check we conclude
that the two different methods we used for GIRAFFE and MIKE targets are
consistent over the entire temperature range.\\
If we move to Na instead, we see a different situation. First of all we considered only stars with
temperature hotter than 4400 K. Again a linear fit gives a slopes
with a significance below 1$\sigma$ and all stars are spread around
the mean value with no sign of any trend. However the two coldest
targets show a Na content much lower than that of their hotter companions.
If we  extrapolate the fit to the temperature of the two coldest stars, we
find a difference with respect to the fit of $\sim$0.50 dex, that is significant at a level of
$\sim$4$\sigma$. This is because they are first generation stars with an
intrinsic low Na, while all the other targets are second generation stars
with high Na. In order to show visually the large difference in Na content of the two
coldest stars with respect to the rest of the sample we plot in
Fig.~\ref{f1d} the spectra of the Na-rich \#D8 (black squares) and the
Na-poor \#S437 (red squares) target around the Na double at 568 nm. Blue lines are the best fitting synthetic
spectra with Na abundances taken from Tab.~\ref{t1}. The green line is a synthetic
spectrum calculated for the atmospheric parameters of \#S437 but with the Na
content of \#D8. It is clear from this comparison that the \#S437 strong Na
under abundance is real and not due, for instance, to the low S/N or to an ill-estimated
stellar effective temperature.

\begin{figure}
 \includegraphics[width=80mm]{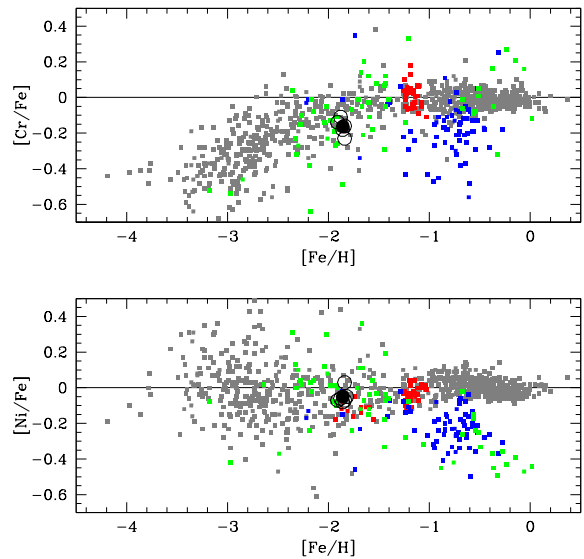}
 \caption{[Cr/Fe] and [Ni/Fe] trends as a function of [Fe/H] for different
   environments (see text). Open black circles indicate MIKE targets, while
   the filled black circle is the mean abundance of NGC~4147.}
 \label{f2}
\end{figure}

\section{Results}

\subsection{Iron-peak elements}

The iron content we obtained from GIRAFFE data is:

$$[Fe/H]_{GIRAFFE}=-1.82\pm0.02$$

while from MIKE data is:

$$[Fe/H]_{MIKE}=-1.86\pm0.01$$

Reported errors are errors on the mean. The difference between the two
datasets is of 0.04 dex, that correspond to 1.8 $\sigma$. The agreement is satisfactory.
Finally we can give a value for the iron content of the cluster that is the
mean of the two datasets:

$$[Fe/H]=-1.84\pm0.02$$

This value well agrees with \citet{Iv09}, that give [Fe/H]=-1.7$\pm$0.1,
and with \citet{Ha96} that give [Fe/H]=-1.80.
The measured iron dispersion in Tab.~\ref{t3} well agrees with the 
dispersion due to measurement errors so we can rule out any intrinsic Fe abundance
spread. 

The chemical abundances for the iron-peak elements Cr and Ni are listed in
Table~\ref{t2}. The value is sub-solar for Cr, while Ni is basically solar-scaled.
Figure \ref{f2} (as well as the following plots) shows the mean (black filled circle) and star by
star (black empty circles) elemental abundances of the cluster compared with a variety of
galactic and extra-galactic objects. We have included values from GGCs 
\citep[red filled squares]{Ca09b,Vi10,Vi11}; Disc and Halo stars
\citep[gray filled squares]{Fu00,Re03,Re06,Ca04,Si04,Ba05,Fr07}
and extra-galactic objects such as Magellanic clouds \citep[blue filled squares]{Po08,Jo06,Mu08,Mu09},  
Draco, Sextans, Ursa Minor and Sagittarius dwarf galaxy and
the ultra-faint dwarf spheroidals Bo\"{o}tes I and Hercules
\citep[green filled squares]{Mo05,Sb07,Sh01,Is14,Ko08}.

Around NGC~4147 metallicity, Galactic and extragalactic environments share the
same iron-peak abundances, with only very few extragalactic stars showing a Cr
depletion. NGC~4147 iron-peak elements agree with both environments and do not
support or disproof an extragalactic origin of this object.

\begin{figure}
 \includegraphics[width=80mm]{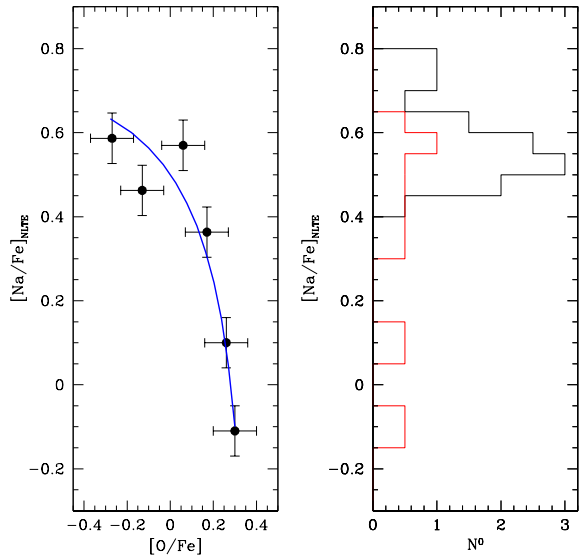}
 \caption{Left panel: [Na/Fe] vs. [O/Fe] as obtained for MIKE targets.
   The cluster shows the typical Na-O anti-correlation of GGCs. The blue line
   is the \citet{Ca09a} diluition model for the cluster. Right panel:
   The Na distribution of GIRAFFE (black) and MIKE (red) targets. See text for more details.}
 \label{f3}
\end{figure}

\begin{figure}
 \includegraphics[width=80mm]{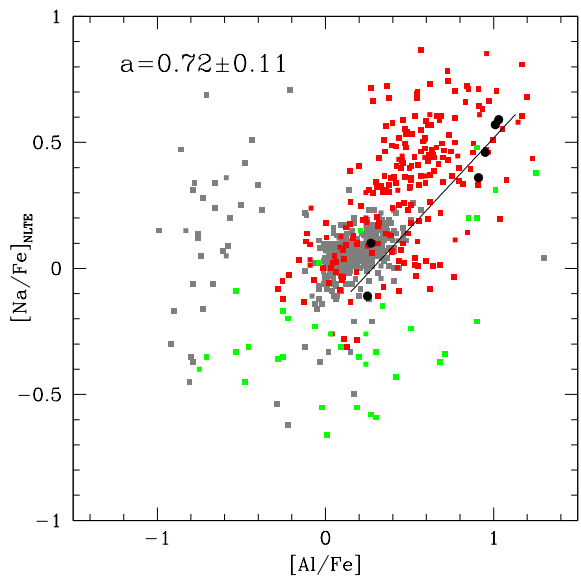}
 \caption{[Na/Fe] vs. [Al/Fe] as obtained for MIKE targets (filled black circles). The cluster shows
   the typical Na-Al correlation of GGCs (red squares). See text for more details.}
 \label{f4}
\end{figure}

\begin{figure}
 \includegraphics[width=80mm]{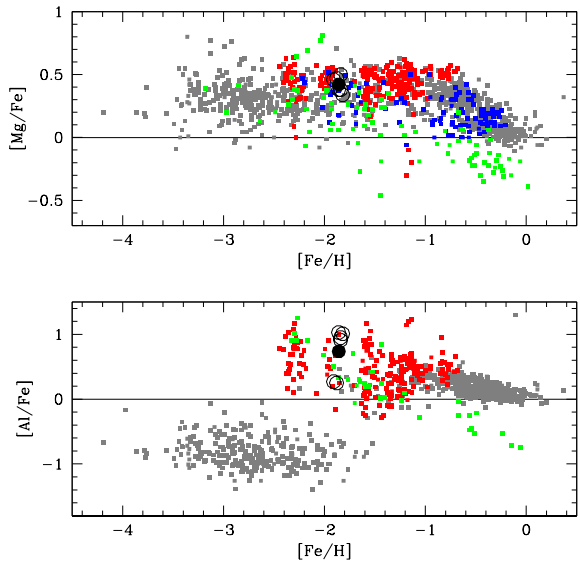}
 \caption{[Mg/Fe] and [Al/Fe] trends as a function of [Fe/H] for different
   environments (see text). Open black circles indicate MIKE targets, while
   the filled black circle is the mean abundance of NGC~4147.}
 \label{f5}
\end{figure}

\subsection{Light elements}

Light elements C, N, O, Na and Al have an observed spread that well exceeds
the observational uncertainties (see Tab.~\ref{t3}). The only exception is Mg that
seems to be homogeneous within the errors. 

Very interesting is the analysis of the Na and O distributions.
In Fig.~\ref{f3} we show the Na-O anticorrelation of MIKE data on the left
panel, while in the right panel we report the Na distribution based on the
GIRAFFE and MIKE data. We fitted the \citet{Ca09a} diluition model
(blue line) to the Na-O anticorrelation. According to this fit the cluster is composed
by a first generation of stars with [Na/Fe]$\sim$+0.0 and [O/Fe]$\sim$+0.3 and
second generation with [Na/Fe]$\sim$+0.5 and [O/Fe]$\sim$-0.2. This is in common
with all but one of the globular clusters studied up to now. The difference
appears when we count the number of
stars for each population. If we combine together the two datasets and
consider all the stars with [Na/Fe]$<$0.30 as first generation, we end up with 
the result that that second generation stars in the cluster represent 
$\sim$90 \%\ of the total, leaving room only for a $\sim$10 \%\ to the first
generation. The poissonian error related with the two estimations is 20 and 10 \%\ respectively.
This result is confirmed by Fig.~\ref{f1}. Here we selected tentatively the HB progeny of the
first generation as red filled circles, and the HB progeny of the
second generation as blue filled circles. If we compare the relative number of
stars, we obtain that second generation stars in the cluster represent 
80$\pm$10 \%\ of the total. However we point out that this number is a lower limits
because we left out blue HB stars evolved off the blue HB zero age sequence,
and at the same time red HB stars suffer a contamination from the field.
So our best estimation is that $\sim$85 \%\ of the stars in NGC~4147
belong to the second generation.
If \citet{Ca11} or \citet{Kh15} are right, we have here a cluster that suffered an extreme loss
of its first generation.

\begin{figure}
\centering
\includegraphics[width=8cm]{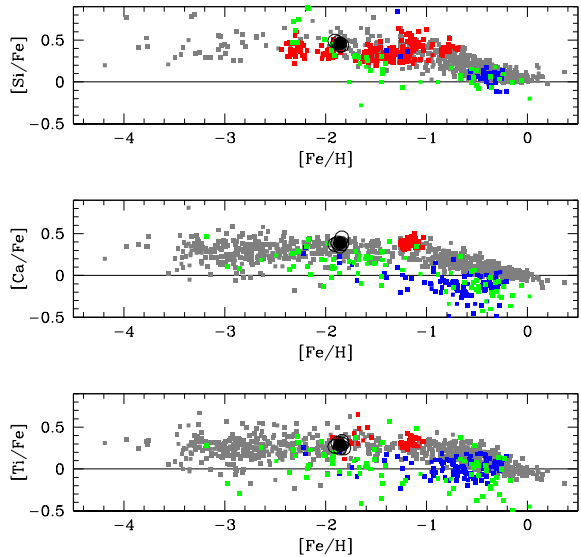}
\caption{[Si/Fe], [Ca/Fe] and [Ti/Fe] trends as a function of [Fe/H] for different
   environments (see text). Open black circles indicate MIKE targets, while
   the filled black circle is the mean abundance of NGC~4147.}
\label{f6}
\end{figure}

In Fig.~\ref{f4} we show the correlation between Na and Al, another elements
usually but not always involved in the light-element spread \citep{Vi11}.
The cluster shows a clear correlation that matches that found in the other
GGCs. In addition the two first generation stars ([Na/Fe]$\sim$0.0,
[Al/Fe]$\sim$0.2) well agrees with the bulk of Galactic objects 
while very few extragalactics targets occupy this region. In the Figure we
report also the slope and its error. The significance of the relation of of $\sim$7$\sigma$.

Finally in Fig.~\ref{f5} we show the Mg and Al vs. Fe trends. NGC~4147
nicely follow the galactic and extragalactic trend as far as Mg is concern.
Al shows a more surprising behavior. If we consider only Milky Way stars,
they follow two separate trends. Almost all stars above [Fe/H]$\sim$-1.7 have
[Al/Fe]$\geq$0, while all stars below [Fe/H]$\sim$-1.7 have [Al/Fe] well below
0 with a mean value of $\sim$-0.8. This result is not totally new,
because \citet[Fig. 5]{Ge06} found the same behavior using [Al/Mg]
vs. [Fe/H] instead of [Al/Fe] vs. [Fe/H] as we do. 
The surprise arrises when  we add GGC and extragalactic dwarfs because in both
classes of objects stars below [Fe/H]$\sim$-1.7 have [Al/Fe]$\geq$0. This
could suggest a different formation history of GCC stars with respect the Halo
field. In any case NGC~4147 follow the GCC trend.

\subsection{ $\alpha$ elements}

The $\alpha$ elements Si, Ca, and Ti are overabundant compared to the Sun. This is a common
feature among almost every GGC as well as among similarly metal-poor
field stars in the Milky Way and in outer galaxies. 
The calcium content we obtained from GIRAFFE data is:

$$[Ca/Fe]_{GIRAFFE}=+0.42\pm0.04$$

while from MIKE data is:

$$[Ca/Fe]_{MIKE}=+0.39\pm0.02$$

Reported errors are errors on the mean. The difference between the two
datasets is of 0.03 dex, within 1$\sigma$. The agreement is satisfactory.
Based on Mg, Si, Ca, and Ti abundances of MIKE stars, we derive for the cluster a mean $\alpha$ element
abundance of:

\begin{center}
$[\alpha/Fe]= +0.39 \pm 0.04$
\end{center}

Figure \ref{f6} shows the $\alpha$-element
abundance of the cluster (MIKE data only), compared with the trend as a function 
of the metallicity for GGCs, disk and halo stars and
extragalactic objects.   
The $\alpha$ elements in NGC~4147 follows the same trend as GGCs and
are fully compatible with Halo field stars. Again
NGC~4147 falls in a region where both Galactic and
extragalactic objects overlap. However if we look carefully at the Ca trend,
we notice that at the metallicity of the cluster the mean Ca abundance for
Halo stars is higher than that for dwarf galaxies. The difference is of
$\sim$0.3 dex. This is true also for Ti. NGC~4147 stars have:

\begin{center}
$[Ca/Fe]= +0.40 \pm 0.02$
\end{center}

and 

\begin{center}
$[Ti/Fe]= +0.28 \pm 0.02$
\end{center}

(reported errors are errors on the mean) that must be compared with the
mean values for the Halo ($[Ca/Fe]\sim0.3$,
$[Ti/Fe]\sim0.3$) and for the Dwarf galaxies ($[Ca/Fe]\sim0.1$,
$[Ti/Fe]\sim0.1$). So, according the its  $\alpha$-element content, NGC~4147
seems more likely to be a genuine Galactic cluster, although some other
clusters like M54 and Arp2, that belong to the Sagittarius galaxy, have a
$\alpha$-element content similar to NGC~4147 ([Ca/Fe]=+0.32 and
[Ca/Fe]=+0.46 respectively,\citealt{Ca10,Mo08}).

\begin{figure}
\centering
\includegraphics[width=8cm]{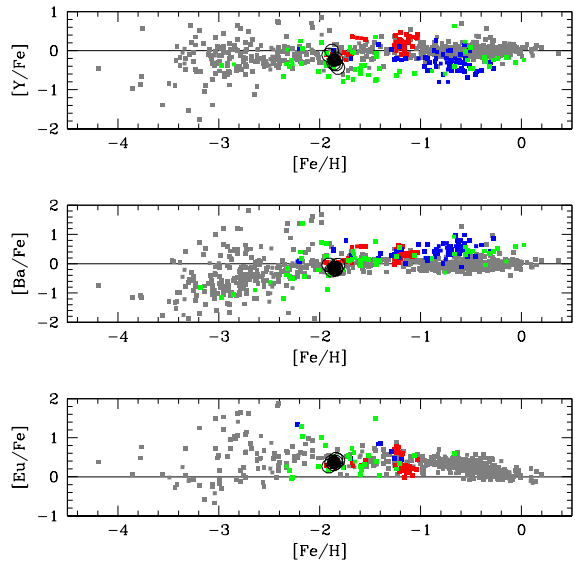}
\caption{[Y/Fe], [Ba/Fe] and [Eu/Fe] trends as a function of [Fe/H] for different
   environments (see text). Open black circles indicate MIKE targets, while
   the filled black circle is the mean abundance of NGC~4147.}
\label{f7}
\end{figure}

\subsection{Heavy Elements}

\begin{figure}
\centering
\includegraphics[width=8cm]{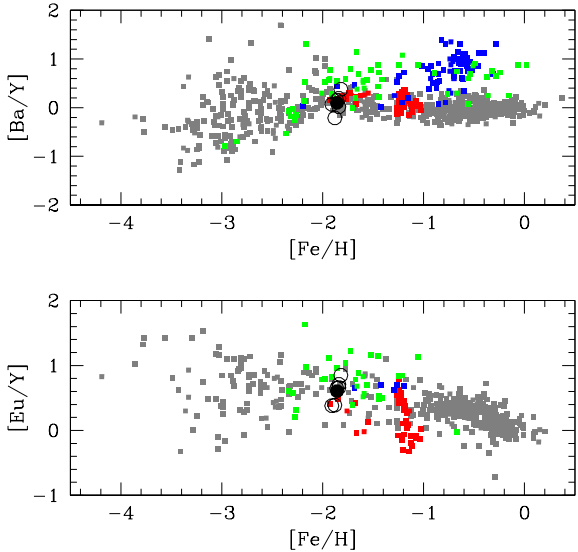}
\caption{[Ba/Y] and [Eu/Y] trends as a function of [Fe/H] for different
   environments (see text). Open black circles indicate MIKE targets, while
   the filled black circle is the mean abundance of NGC~4147.}
\label{f8}
\end{figure}

While light-element variations are well-known in GCs, intrinsic dispersion
among heavier elements is less common. Most of the heavier elements (Z $>$ 30)
are produced either by slow or rapid neutron-capture reactions (the so-called
s and r processes). s-process happens in a different
physical condition with respect to r-process and are thus likely to
happen in different astrophysical sites.  We measured the abundances of the
neutron-capture elements: Y, Ba, and Eu. Y and Ba are mainly produced by
the s-process at solar metallicity, while Eu is produced almost exclusively
in the r-process. The mean Y and Ba abundance ratios are slightly sub-solar,
while Eu is heavily super-solar.
Figure \ref{f7} shows the mean heavy-element
ratios of the cluster (MIKE data only), compared with the trend as a function 
of the metallicity of GGCs, disk and halo stars and
extragalactic objects. All three elements agree with Milky Way Halo as well
as with extragalactic environment.
Figure \ref{f8} shows the mean [Ba/Y] and [Eu/Y] ratios, where again
NGC~4147 agrees both with the Galactic and the extragalactic trends

\begin{figure}
\centering
\includegraphics[width=8cm]{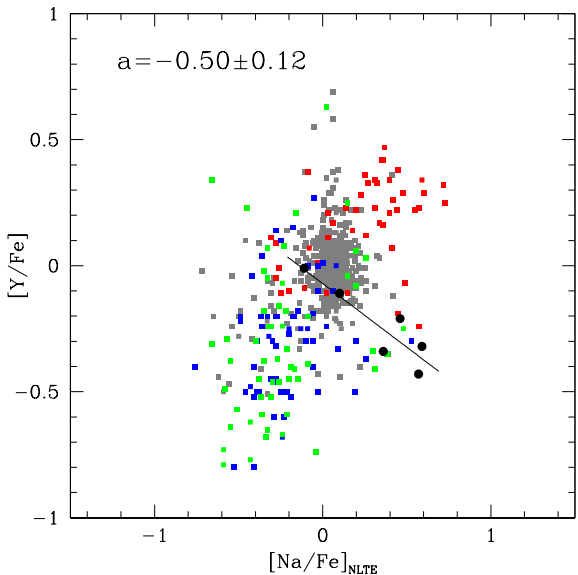}
\caption{[Y/Fe] vs. [Na/Fe] as obtained for MIKE targets (filled black circles). The cluster shows
   a possible Y-Na anti-correlation. See text for more details.}
\label{f9}
\end{figure}

Ba and Eu show an observed spread that is comparable to
that expected from the errors (see Table~\ref{t3}), while for Y the observed spread
is significantly larger. To check this behavior, we show in Figure~\ref{f9} 
the abundance ratios of [Y/Fe] as a function of [Na/Fe].
The two first generation stars ([Na/Fe]$\sim$0.0) well agrees with the bulk of
Galactic objects while most of the extragalactic stars have lower Na and Y
content. This support our identification of NGC~4147 as a Galactic cluster.
What is more interesting is the anticorrelation between Y and Na
with Y that decreases while Na increases. In the plot we
reported the slope a of the linear fit to NGC~4147 stars (black line) together
with its error. The significance of the fit is larger than 4$\sigma$. 

Only a handful of metal-poor globular clusters show a potential star-to-star
dispersion in neutron-capture elements \citep{Ka13,Wo13}.
\citet{Ma09} found a wide range of abundances values for
s-process elements, Y, Zr and Ba, in M22. They also identified a
bimodality among these elements. However none of the elements show a
correlation with Na, O and Al.

Our result implies that the second Na-rich generation was formed by
material where Y was destroyed. This is at odd with current models that
describe the multiple population phenomenon in GCs, which postulate AGB stars
as major polluters \citep{Ve09}. In fact AGB stars also produce light
s-process elements like Y \citep{Cr15}. The amount of light s-element produced
depends a lot on the mass of the AGB star, but in no case a Y destruction is
predicted. For this reason we checked for possible blending not considered in our
line-list. For example a blend with a CN or CH line could mimic a [Y/Fe] spread
even if Y is constant just because C and N vary. However the result of this
check was negative. We underline also that the fact that the two Y-rith first
genaration stars are also the two coldest objects in our sample. We have no
first generation stars in the temperature range of the second generation
targets, so we cannot totally roule out a possible temperature effect on our Y
abundance determination. In spite of that we leave open the possibility that the
[Y/Fe] vs. [Na/Fe] trend we found is real waiting for future studies.

\section{Summary}

In this paper we present the first detailed chemical abundances of 15 elements
in 6 red giant radial velocity members of NGC~4147 observed using the high resolution MIKE
spectrograph, mounted at the Magellan-Clay telescope, and Fe and Na abundances
in 12 red giant radial velocity members observed using the medium
resolution GIRAFFE spectrograph, installed at the VLT-UT2 telescope.
Chemical abundances have been computed using plane-parallel atmospheric-models
and LTE approximation. Equivalent width method has been used when possible. Otherwise we applied
the spectrum-synthesis method. We obtained the following results:

\begin{itemize}
 \item We found a mean metallicity of [Fe/H]=-1.84$\pm$0.02, that well agree
       with previous literature data. As far as other iron-peak elements are
       concerned, Cr is sub-solar while Ni is solar-scaled, in agreement with
       Halo and dwarf galaxies environment at the cluster metallicity
 \item NGC~4147 shows the typical Na-O anticorrelation common to almost all
       the other GGCs. However the cluster contains only $\sim$15\% of first
       generation stars. This implies that it suffered a severe mass loss,
       maybe the most extreme among all the GGCs. Na is also correlated with
       Al as found in many GGCs.
       Mg follows the Galactic trend, while Al is much more enhanced with
       respect to Halo stars sharing this behaviors with all the other GGCs.
 \item NGC~4147 has the typical $\alpha$-enhancement of the Halo. 
       Its Ca and Ti abundances agree better with the Halo than with the mean 
       Ca and Ti content of extragalactic environments at the cluster
       metallicity. This is true also if we consider the behavior of
       [Na/Fe] vs. [Al/Fe] and [Y/Fe] vs. [Na/Fe].
 \item Heavy elements Y, Ba and Eu have mean abundances that match those of
       the Milky Way and of extragalactic environments. While Ba and Eu are
       homogeneous, Y possibly shows a spread and an anticorrelation with Na.
 
\end{itemize}

\section*{Acknowledgments}

SV and CMB acknowledge the support provided by Fondecyt Regular 1130721 and
1150060, respectively. PA acknowledges the support provided by the Chinese
Accademy of Sciences President's International Fellowship Initiative, Grant
N. 2014FFJB0018, and by the Programa de Astronomia, Fondo China-Conicyt 2Conv. CAS15020.

\bsp

\label{lastpage}

\end{document}